\documentclass[]{spie}  

\usepackage[]{graphicx}

\title{Suppressed Andreev Reflection at the Normal-Metal / Heavy-Fermion Superconductor CeCoIn$_5$ Interface} 

\author{Wan Kyu Park\supit{a}, Laura H. Greene\supit{a}, John L. Sarrao\supit{b}, and Joe D. Thompson\supit{b}
\skiplinehalf
\supit{a}Department of Physics and the Frederick Seitz Materials Research
Laboratory, University of Illinois at Urbana-Champaign, Urbana, IL 61801, USA; \\
\supit{b}Los Alamos National Laboratory, Los Alamos, NM 87545, USA
}

\authorinfo{Further author information: (Send correspondence to Wan Kyu Park or Laura H. Greene)\\ Wan Kyu Park: E-mail: wkpark@uiuc.edu, Telephone: 1 217 265 5010\\ Laura H. Greene: E-mail: lhg@uiuc.edu, Telephone: 1 217 333 7315}

\begin{document} 
\maketitle 

\begin{abstract}
In order to probe the order parameter symmetry of the heavy-fermion superconductor (HFS) CeCoIn$_5$, we employ point-contact spectroscopy, where dynamic conductance spectra are taken from a nano-scale junction between a normal-metal (N) Au tip and a single crystal of CeCoIn$_5$. The point-contact junction (PCJ) is formed on a single crystal surface with two crystallographic orientations, (001) and (110).
Our conductance spectra, reproducibly obtained over wide ranges of temperature, constitute the cleanest data sets ever reported for HFSs. The point contacts are shown to be in the Sharvin limit, ensuring spectroscopic nature of the measured data. A signature for the emerging heavy-fermion liquid is evidenced by the development of the asymmetry in the background conductance, starting at $T^*$ ($\sim$ 45 K) and increasing with decreasing temperature down to $T_c$ (2.3 K). Below $T_c$, an enhancement of the sub-gap conductance arising from Andreev reflection is observed, with the magnitude of $\sim$ 13.3 \% and $\sim$ 11.8 \% for the (001) and the (110) PCJ, respectively.
These values are an order of magnitude smaller than those observed in conventional superconductors, but consistent with those in other HFSs. Our zero-bias conductance data for the (001) PCJ are best fit with the extended Blonder-Tinkham-Klapwijk model using the $d$-wave order parameter. The fit to the full conductance curve of the (001) PCJ at 400 mK indicates the strong coupling nature ($2\Delta/k_{B}T_c = 4.64$). However, our observed suppression of both the Andreev reflection signal and the energy gap indicates the failure of existing models. We provide possible directions for theoretical formulations of the electronic transport across an N/HFS interface in general, and the Au/CeCoIn$_5$ interface in particular. Several qualitative features observed in the (110) PCJ provide the first clear spectroscopic evidence for the $d_{x^2-y^2}$ symmetry of the superconducting order parameter in CeCoIn$_5$.
\end{abstract}


\section{INTRODUCTION}
\label{sect:intro}

Probing the order parameter symmetry is of primary importance in investigating the pairing mechanism of unconventional superconductors.\cite{bennemann03} Phase sensitive experiments such as Josephson junctions and SQUIDs\cite{vanharlingen95} and single particle tunneling spectroscopy\cite{covington97} have been playing important roles in manifesting the $d$-wave symmetry of hole-doped cuprate high-$T_c$ superconductors (HTSs). Heavy-fermion superconductors (HFSs), another class of unconventional superconductors, have been studied mostly by point-contact spectroscopy\cite{naidyuk98,lohneysen96,naidyuk05} (PCS), partly because these materials are difficult to grow as thin films for tunnel junctions.  

Of the HFSs discovered, the relatively new family, CeMIn$_5$ (M = Co, Rh, Ir), have been of great research interest since they exhibit rich physical phenomena (Ref.~\citenum{thompson03} and references therein) such as quantum phase transitions\cite{sidorov02,paglione03} and the Fulde-Ferrell-Larkin-Ovchinikov (FFLO) phase transition.\cite{radovan03,bianchi03} There have been many reports on the symmetry of the superconducting order parameter in CeMIn$_5$. In particular, there exist substantial evidences for line nodes in the order parameter of CeCoIn$_5$.\cite{movshovich01,izawa01,eskildsen03,aoki04} The $d$-wave pairing symmetry is a most likely candidate but not conclusive yet since {\it definite spectroscopic proof is still lacking}.
Goll \textit{et al}.\cite{goll03} and Rourke \textit{et al}.\cite{rourke05} have reported PCS data on CeCoIn$_5$, claiming that the superconducting order parameter has an unconventional symmetry\cite{goll03,rourke05} and even multiple components.\cite{rourke05} However, it can be shown that these data\cite{rourke05} may not reflect intrinsic properties of CeCoIn$_5$.\cite{sheet05,park05comm} Furthermore, it remains controversial whether line nodes are located along the (100)-axis\cite{izawa01} or the (110) direction.\cite{aoki04} Addressing this issue is important for the clarification of either $d_{x^2-y^2}$-wave\cite{izawa01,eskildsen03} or $d_{xy}$-wave\cite{aoki04} symmetry.

PCS, due to its simplicity and versatility as a spectroscopic tool, has been widely adopted for the investigation of both conventional and unconventional superconductors including HFSs\cite{lohneysen96,naidyuk98,naidyuk05}. In general, it can provide information on the density of states and the gap energy ($\Delta$) through conductance data taken from a nano-scale junction between a normal-metal (N) and a superconductor (S). 
Blonder, Tinkham, and Klapwijk\cite{blonder82,blonder82a} (BTK) formulated a theoretical model for the electronic transport across an N/S interface. The BTK theory provides clear descriptions on the transitional behavior from a metallic to a tunnel junction using the effective barrier strength, $Z_{eff}$, as a single parameter. Therefore, it has been playing a crucial role for the analysis of PCS data.\cite{naidyuk05}

If a quasi-particle (QP) is injected with energy lower than $\Delta$ from the normal-metal toward the superconductor in an N/S contact, it cannot enter the superconductor as a single particle since there are not available single particle states below $\Delta$ in the superconductor.
This QP can transport to a superconductor by being retro-reflected as a quasi-hole and forming a Cooper pair with another electron. This Andreev reflection\cite{andreev64,deutscher05} (AR) is a quantum mechanical scattering from a superconducting pair potential, conserving energy, momentum, spin, and charge.
In a pure metallic N/S contact with $Z_{eff}=0$, the zero-bias conductance (ZBC) is predicted
to be doubled compared to that at high-bias ($V \gg \Delta/e$),\cite{blonder82,blonder82a} so that the ZBC is enhanced by 100 \%. A substantial enhancement of the ZBC due to AR was observed in point contacts containing conventional superconductors.\cite{naidyuk05,blonder82a,blonder83}

According to the BTK theory,\cite{blonder82a,blonder83} the Fermi velocity mismatch acts as an effective barrier, thereby reducing the probability for AR. The effective barrier strength is given by $Z_{eff}=[(1-r)^2/4r +Z_0^2]^{1/2}$, where $r \equiv v_{FN}/v_{FS}$, the ratio of the Fermi velocities in the electrodes (note $Z_{eff}$ remains invariant for $r\rightarrow1/r$), and $Z_0$ is the barrier strength due to an insulating layer.
In mesoscopic semiconductor-superconductor junctions, results reported for Si-, GaAs-, InGaAs-, and InAs-based junctions with Nb counter-electrodes\cite{smpcs} could be accounted for using this formula.
However, since it is not possible to separate the effects of an impurity- or disorder-induced barrier ($Z_0$) at the interface from that of the Fermi surface mismatch in these systems, the accuracy of
$Z_{eff}$ remains inconclusive.
An N/HFS point contact, due to the large disparity of the Fermi velocities, is expected to behave as a tunnel junction ($Z_{eff}>5$).
However, an enhancement of the sub-gap conductance (ESGC) due to AR has been commonly observed in many N/HFS point contacts, albeit suppressed in magnitude.\cite{goll93,wilde94,goll95,wilde96,naidyuk96PhysicaB,obermair98}
Deutscher and Nozi{\'e}res \cite{deutscher94} addressed this inconsistency between the theory and the experiments by proposing that the Fermi velocities entering in the ratio \textit{r} are
without a mass enhancement factor.

Here, we report conductance spectra taken from point contacts on CeCoIn$_5$. We do observe an AR-induced enhancement of the conductance but with heavily reduced magnitudes, implying that our results are consistent with reports on other HFSs.\cite{goll93,wilde94,goll95,wilde96,naidyuk96PhysicaB,obermair98}
Furthermore, the dependence of our data on the crystallographic orientations provides the first clear spectroscopic evidence for the $d$-wave symmetry of the superconducting order parameter.

\section{EXPERIMENTS}
\label{sect:exp}

\begin{figure}[t]
\begin{center}
\includegraphics{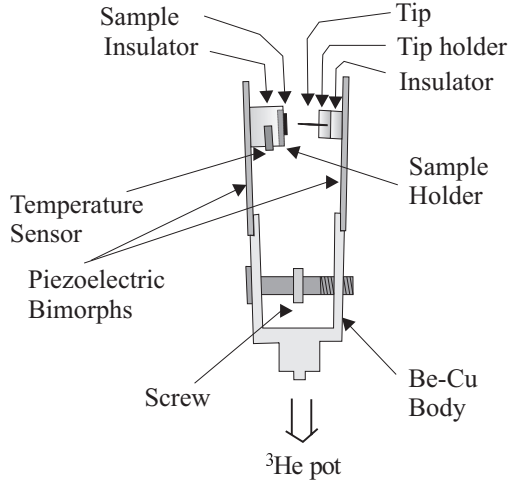}
\end{center}
\caption{\label{fig:CATrig}
A schematic drawing of the point-contact spectroscopy rig (not to scale). A point contact is formed by bringing a tip and a sample into contact using a fine screw. It is kept stable by adjusting the $dc$ voltage to the piezoelectric bimorphs while being cooled down. The main body is made of Be-Cu to ensure a reversible movement when adjusted by the screw.}
\end{figure}

We have developed a PCS technique based on a combination of mechanical and piezoelectric mechanisms for making point contacts.\cite{park05rsi} A schematic drawing of our PCS rig is displayed in Fig.~\ref{fig:CATrig}. A finely polished Au tip is prepared by etching a Au wire in a concentrated hydrochloric acid with a $dc$ pulse applied between it and a Pt counter-electrode. After the etching process, the surface is examined using an optical microscope and a scanning electron microscope (SEM). Only Au tips with smooth and clean surfaces are used. This is important to avoid the formation of non-ballistic and/or multiple contacts and degraded spectroscopic features.

High quality CeCoIn$_5$ single crystals are grown using excess In flux \cite{petrovic01}. Considering the tetragonal structure of CeCoIn$_5$, three kinds of point contacts are prepared with the surface having a different crystallographic orientation, named as (001), (110), and (100) point-contact junctions (PCJs). The (001) PCJs are formed on the largest surface of as-grown crystals since CeCoIn$_5$ single crystal is known to grow along the $c$-axis. Samples for (110) and (100) PCJs are prepared by embedding single crystals into a mold of a low temperature epoxy and cutting them such that the exposed surface is normal to the (110) and the (100) direction, respectively. These pieces are polished using alumina and/or diamond lapping films and silica colloidal suspensions down to 25 nm particle size. The polished surface, examined using an optical microscope and an SEM, looks mirror-like shiny and smooth. The actual crystallographic orientation of the sample is checked by the X-ray diffraction.

After a Au tip and a single crystal of CeCoIn$_5$ are mounted onto the PCS rig, they are brought into contact by adjusting a fine screw at room temperature. Then, the PCS rig is attached to the $^3$He cryostat. The point contact is kept stable by adjusting the $dc$ driving voltage to the piezoelectric bimorphs during the cool-down process. The lowest achievable temperature is 300 mK and the maximum magnetic field is 12 Tesla.
The dynamic conductance spectra of a PCJ are taken using the standard four probe lock-in technique as a function of bias voltage, temperature, and magnetic field. We report conductance spectra obtained from (001) PCJs and provide extensive analysis and discussion.\cite{park04} In addition, conductance data of (110) PCJs are presented with qualitative discussions. Results on (100) PCJs are not reported here, although some preliminary data have been obtained. 

\section{RESULTS AND DISCUSSION}
\label{sect:results}

\subsection{Conductance of (001) Point-Contact Junction}

\begin{figure}[b]
\begin{center}
\includegraphics{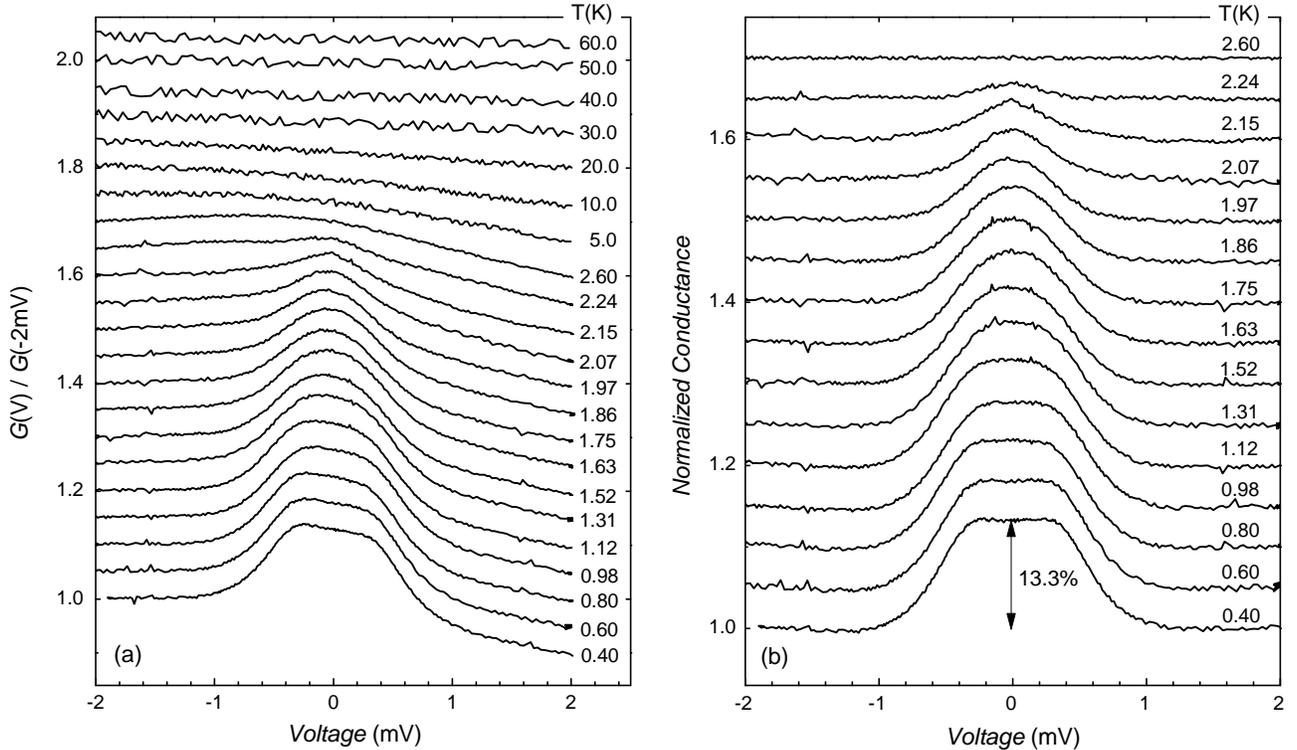}
\end{center}
\caption{\label{fig:001temp-dep} Conductance vs. voltage data of a Au/CeCoIn$_5$ (001) point-contact junction. Curves are shifted vertically by 0.05 for clarity. (a) Conductance data are normalized to the conductance at --2 mV. An asymmetry is seen to develop at $\sim$ 45 K, increasing with decreasing temperature, but remains constant below $T_c$. (b) Conductance curves are symmetrized by dividing the data in (a) with that at 2.60 K. The enhancement of conductance near zero-bias, 13.3 \% at 400 mK, is due to Andreev reflection.}
\end{figure}

The $dI/dV$ vs. $V$ spectra obtained from a (001) PCJ are displayed in Fig.~\ref{fig:001temp-dep}(a), normalized to the conductance at --2 mV. The PCJ was stable over a wide temperature range from 60 K to 400 mK. In the normal state, an asymmetry in the background conductance is seen to develop, starting at $\sim$ 45 K. We attribute this to the \textit{emergence of a coherent heavy-fermion liquid}.\cite{nakatsuji04} This asymmetry is enhanced with decreasing temperature down to 2.6 K, below which it remains almost constant. This behavior is consistent with the observation that the relative weight of a coherent phase saturates below $\sim$ 2 K.\cite{nakatsuji04} These coincident behaviors of the background conductance with the emergence of the two fluids may provide important clues to our understanding of the electronic transport in CeCoIn$_5$. However, more detailed theoretical investigations are required.

The constancy of the background conductance below $T_c$ enables us to perform quantitative analysis. That is, we can obtain symmetrized conductance spectra by factoring out the asymmetric part of the conductance data by that taken at 2.6 K, as shown in Fig.~\ref{fig:001temp-dep}(b). Below $T_c$, it is observed that the conductance near zero-bias is enhanced. This ESGC increases with decreasing temperature, showing a flat region at lower temperatures. We attribute this ESGC to AR at the Au/CeCoIn$_5$ interface. At higher temperatures, this region becomes rounded due to the thermal broadening of the Fermi distribution function. As indicated in Fig.~\ref{fig:001temp-dep}(b), the ZBC at the lowest temperature is enhanced by 13.3 \%, which is much smaller than those in N/conventional superconductor point contacts with small $Z_{eff}$ but consistent with the results for other N/HFS point contacts.\cite{goll93,goll95,obermair98} We take this reduced ESGC as a ubiquitous penomenon in N/HFS point contacts, implying that there might be intrinsic mechanisms to cause AR to be suppressed at an N/HFS interface. 

In PCS measurements, it is important to ensure that conductance spectra are taken from a ballistic point contact to guarantee the {\it spectroscopic} nature of the data. Since it is not feasible to measure the size of a point contact directly, typically it is estimated using the measured resistance at a high bias ($V \gg \Delta/e$), and Wexler's formula\cite{wexler66},
\begin{equation}
R_0 = \frac{4\rho l}{3\pi a^2}\{ 1 + \frac{3\pi}{8K} \gamma(K) \},
\label{eq:wexler}
\end{equation}
where $R_0$ is the resistance of the contact, $K \equiv l/a$, $a$ the contact radius, $\gamma(K)$ a smooth function of $K$, $\rho$ and $l$ the resistivity and the electronic mean free path (EMFP) in electrodes.
In the Sharvin\cite{sharvin65} or ballistic limit ($K \gg 1$), $\gamma(K)\to 0.694$,
and $R_0 = R_S = \frac{4\rho l}{3\pi a^2}$.
In the Maxwell\cite{holm67} or diffusive limit ($K \to 0$), $\gamma(K) \to 1$,
and $R_0 = R_M = \frac{\rho}{2a}$.
Since the low-temperature resistivity of Au is negligible compared to the reported value of CeCoIn$_5$,\cite{movshovich01} the contact size is estimated using only the latter. We also note that, for a point contact with non-zero $Z_{eff}$, $R_0$ in Eq.~(\ref{eq:wexler}) is related to the measured resistance, $R_N$, through the relation $R_N=R_0(1+Z_{eff})$.\cite{blonder83}
$R_N \sim 1.1\ \Omega$ and $Z_{eff} \sim 0.365$ (obtained through the analysis in a later section) are used. Since $K$ is not known in advance, we estimate the contact size in both ballistic and diffusive limits, obtaining an upper limit for the contact size, $2a \le 460$ \AA.

Movshovich {\it et al}. reported an estimation of the elastic EMFP to be 810 \AA \ at $T_c$ from the thermal conductivity data on CeCoIn$_5$.\cite{movshovich01} Therefore, our measured point contact is much smaller than the EMFP at $T=T_c$, indicating that our conductance spectra were taken from a point contact in the Sharvin or ballistic limit.
An estimation of the EMFP can be extended below $T_c$ using the same data\cite{movshovich01}
and the following thermodynamic relations for low-energy QPs in a $d$-wave superconductor.\cite{hussey02}
\begin{equation} 
\kappa/T \propto \rho_n \tau,
\ \rho_n \propto T,
\ \tau = l/v_F,
\ \textrm{therefore},
\ \ l \propto \kappa/T^2,
\label{eq:thdyrelation}
\end{equation}
where $\kappa$ is the thermal conductivity, $T$ the temperature, $\rho_n$ the density of normal QPs, $\tau$ the QP lifetime, and $v_F$ the Fermi velocity.
It is found that $l$ increases exponentially with decreasing temperature, ranging 4\ -- 5 $\mu$m at 400 mK, nearly two orders of magnitude larger than the contact size. 
We also estimate the inelastic EMFP based on the microwave conductivity data,\cite{ormeno02} obtaining a lower limit of $\sim$ 6500 \AA \ at 400 mK. 
Therefore, we can ensure that \textit{the measured contact is truly in the ballistic or Sharvin limit}
at low temperatures, even if we take into account the possibility of reduction of the EMFP in a point contact due to possible changes of the materials properties at the junction area from those of the bulk materials. These estimated values of the EMFP also show that CeCoIn$_5$ is in the extreme clean limit ($l \gg \xi$, where $\xi$ is the coherence length), which together with the Pauli-limited upper critical field has been reported to be essential for the observation of the long-standing FFLO phase transition\cite{radovan03,bianchi03} in this material.
It is also clear that the arguments proposed by Gloos \textit{et al}.,\cite{gloos96} attributing the suppressed ESGC to the non-ballistic nature of the contact, are not valid for our PCS measurements.

As mentioned before, PCS data can provide information on the electronic density of states and the order parameter symmetry of a superconductor through analysis based on the BTK theory. Since the presence of line nodes in CeCoIn$_5$ has been reported,\cite{movshovich01,izawa01,eskildsen03,aoki04} we choose the extended version of the BTK theory to a $d$-wave superconductor by Tanaka and Kashiwaya,\cite{tanaka95,kashiwaya96,kashiwaya00} hereafter called as the EBTK model. According to this model, the conductance is given by an integration of the conductance kernel over appropriate energy and momentum spaces.
We have calculated the conductance of an N/$d$-wave superconductor junction as a function of bias voltage for three different cases, that is, the junction normal along the $c$-axis, lobe, and nodal directions of the superconductor.
For a (001) PCJ, the conductance kernel is simplified due to its symmetry with respect to electronic trajectories. If the direction of QP momentum is strictly along the $c$-axis, which is the case of the extreme tunneling limit, no AR can occur. In an N/S point contact, typically $Z_{eff}$ is small. Therefore, if we follow the line of reasoning for the tunneling cone (the cone angle is inversely proportional to $Z_{eff}$), the conductance is obtained by the integration over the full half of the momentum space assuming that the momentum of the QP is distributed over this space:
\begin{equation}
\frac{dI}{dV}(V)=
\frac{\int_{0}^{2\pi}d\phi\int_{0}^{\frac{\pi}{2}}
d\theta\int_{-\infty}^{\infty}dE
\frac{df(E-eV)}{dV}
\sigma_S(E,\phi)\sin(2\theta)}
{\int_{0}^{2\pi}d\phi\int_{0}^{\frac{\pi}{2}}d\theta
\sigma_N \sin(2\theta)},
\label{eq:kernel}
\end{equation}
where $f$ is the Fermi distribution function,
$\sigma_S(E,\phi) = \frac{1+|\Lambda|^2+Z^2(1-|\Lambda|^4)} {|1+Z^2(1-\Lambda^2)|^2},
\Lambda = \frac{E-\sqrt{E^2-|\Delta|^2}}{|\Delta|}$,
$\sigma_N = \frac{1}{1+Z^2}$, and $Z = \frac{Z_{eff}}{\cos \theta}$.
For the $d$-wave symmetry, $\Delta(T, \phi) = \Delta(T) \cos 2\phi$.
In order to incorporate the effect of the QP lifetime broadening, we replace $E = E' -i\Gamma$, where $\Gamma = \hbar / \tau$ is the QP scattering rate, and take the real part of the kernel.\cite{dynes78}

\begin{figure}[t]
\begin{center}
\includegraphics{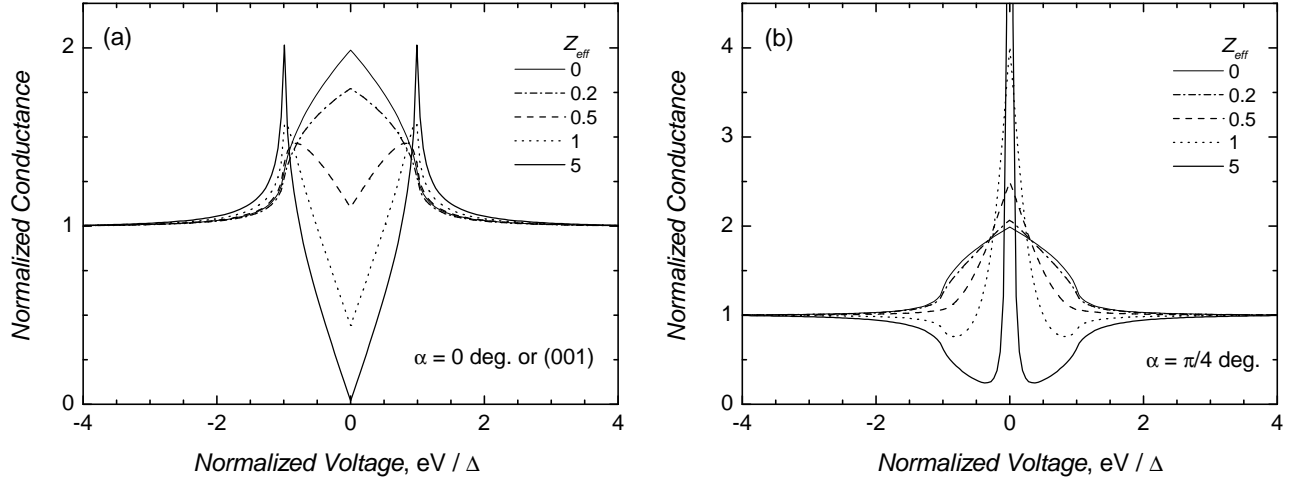}
\end{center}
\caption{\label{fig:eBTK-calc} Normalized conductance of a normal-metal/$d$-wave superconductor junction calculated using the extended BTK model with various effective barrier strengths, $Z_{eff}$. $\alpha$ is the angle between the junction normal and the lobe direction of the $d$-wave order parameter. (a) The junction normal is along the lobe or $c$-axis direction. With increasing $Z_{eff}$, a transitional behavior from Andreev reflection to tunneling is clearly seen, together with a dip in the zero-bias conductance for intermediate $Z_{eff}$. (b) The junction normal is along the nodal direction. The zero-bias conductance peak becomes narrower and stronger with increasing $Z_{eff}$.}
\end{figure}

\begin{figure}[t]
\begin{center}
\includegraphics{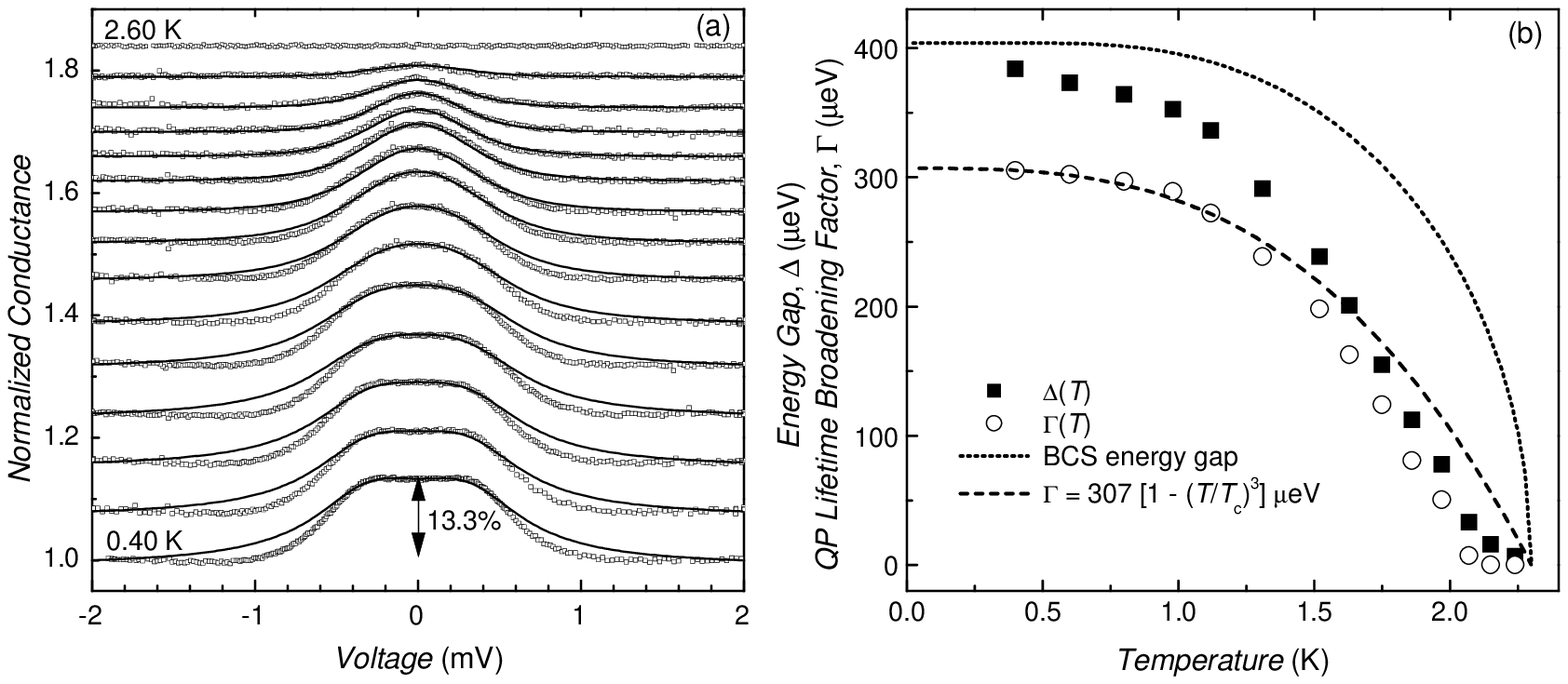}
\end{center}
\caption{(a) The best fit curves (solid lines) to the normalized conductance spectra (open squares) using the $s$-wave BTK model. Calculated curves and data are shifted for clarity. The temperature is, from the bottom to the top, 0.40, 0.60, 0.80, 0.98, 1.12, 1.31, 1.52, 1.63, 1.75, 1.86, 1.97, 2.07, 2.15, 2.24, and 2.60 K, respectively.
(b) Parameters used for the $s$-wave BTK fitting, $\Delta$ (filled squares) and $\Gamma$ (open circles), together with the BCS energy gap (solid line) and $\Gamma$ (dashed line) used in the fitting of the zero-bias conductance data in Fig.~\ref{fig:001-400mK-n-zbc-fit}(b). $\Delta$ shows a reasonable temperature dependence, whereas $\Gamma$ decreases with increasing temperature, which is unphysical.}
\label{fig:001s-wavefit}
\end{figure}

Figure~\ref{fig:eBTK-calc} shows conductance curves calculated for zero temperature and various values of $Z_{eff}$. Here, $\Gamma$ is set to zero. $\alpha$ is the angle between the junction normal and the lobe direction of the $d$-wave order parameter. In Fig.~\ref{fig:eBTK-calc}(a), it is seen that the conductance curves are identical for junctions with the normal along the $c$-axis and the lobe direction because of the reflection symmetry of the order parameter. The transitional behavior from AR to tunneling with increasing $Z_{eff}$ is clearly demonstrated. AR do appear for a junction with small $Z_{eff}$, forming a conductance dip at the zero-bias with increasing $Z_{eff}$. On the other hand, a junction with the normal along the nodal direction exhibits strikingly different conductance features, as displayed in Fig.~\ref{fig:eBTK-calc}(b). For $Z_{eff} = 0$, the conductance curve looks the same as in Fig.~\ref{fig:eBTK-calc}(a), i.e., the usual AR-induced conductance is obtained. However, the conductance curve becomes narrower and sharper with increasing $Z_{eff}$, forming a zero-bias conductance peak (ZBCP) instead of a dip. The ZBCP in a tunnel junction made of a hole-doped HTS, a $d$-wave superconductor, has been frequently observed\cite{covington97} and well understood theoretically.\cite{fogelstrom97} It originates from the constructive interference between an incoming electron and an Andreev-reflected hole due to the phase shift of $\pi$ in the order parameter at a surface of $d$-wave superconductor whose normal is along the nodal direction. As a result, QPs form bound states at zero energy, called Andreev Bound States (ABSs). As demonstrated in Fig.~\ref{fig:eBTK-calc}(b), the ABS-induced ZBCP is smeared out with decreasing $Z_{eff}$,\cite{kashiwaya96} so that the conductance curve for $Z_{eff}=0$ is the same as in Fig.~\ref{fig:eBTK-calc}(a).

\begin{figure}[t]
\begin{center}
\includegraphics{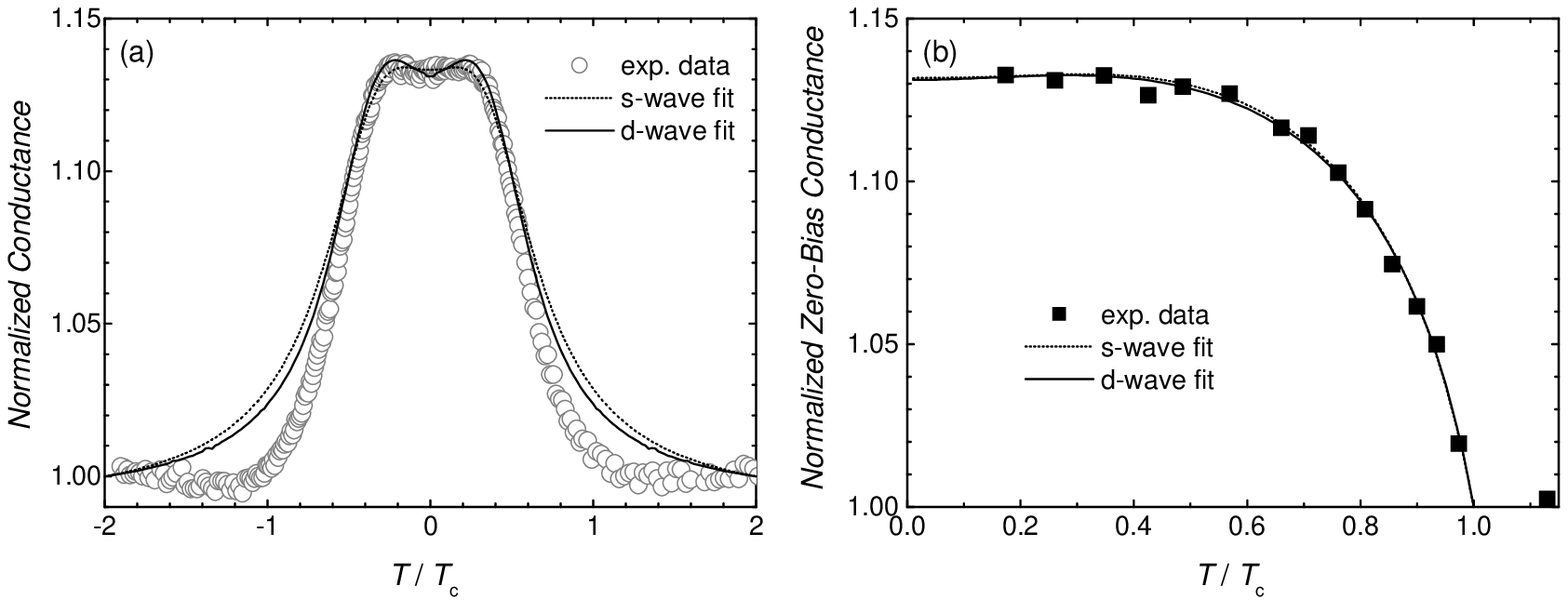}
\end{center}
\caption{(a) The best fit curve (solid line) to the conductance data (open circles) at 400 mK using the $d$-wave BTK model, together with the $s$-wave fit curve (dotted line). The $d$-wave fit gives rise to a slightly better result, reproducing the dip-peak structure near zero-bias, but the deviation is still substantial.
(b) The zero-bias conductance vs. temperature data (filled squares) and the best fit curves using the $s$-wave (dotted line) and the $d$-wave (solid line) BTK models. Both fits are equally satisfactory, however, the temperature dependence of $\Gamma$ is unphysical for the $s$-wave fit (see Fig.~\ref{fig:001s-wavefit}(b)), whereas acceptable for the $d$-wave fit ($\Gamma=218\ \mu \textrm{e}V$).}
\label{fig:001-400mK-n-zbc-fit}
\end{figure}

Fitting to the data is performed by numerical integration of Eq.~(\ref{eq:kernel}) using
$Z_{eff}$, $\Delta$, and $\Gamma$ as parameters. Comparison of the experimental data at 400 mK in Fig.~\ref{fig:001temp-dep}(b) to the calculation in Fig.~\ref{fig:eBTK-calc}(a) implies that $Z_{eff}$ may have a value between 0.2 and 0.5.
Although there exist many reports on line nodes in CeCoIn$_5$, \cite{movshovich01,izawa01,eskildsen03,aoki04} $s$-wave fitting is also performed for completeness. 
Since $R_N$ remains almost constant below $T_c$, it is reasonable to set $Z_{eff}$ to a constant. The optimum value we obtain for $Z_{eff}$ is 0.346.
We also find that varying $Z_{eff}$ as a function of temperature does not result in any better fits
for either the $s$- or the $d$-wave BTK model. The best fit curves using the $s$-wave BTK model are displayed as solid lines in Fig.~\ref{fig:001s-wavefit}(a). The flat region near zero-bias in the low temperature data are reproduced well by adjusting $\Delta$ and $\Gamma$. The calculated curve fits the data well near $T_c$, whereas noticeable deviations are seen around the gap edge, these deviations increasing with decreasing temperature. This deviation is sometimes attributed to a local Joule heating effect due to the contact being in the thermal regime.\cite{naidyuk98,gloos00} However, this cannot be the origin in our point contact spectra since the contact is shown to be in the ballistic limit over the measured temperature range.
The best fit values of $\Delta$ and $\Gamma$ are plotted in Fig.~\ref{fig:001s-wavefit}(b) as a
function of temperature. The gap energy extrapolated to zero temperature, $\Delta(0) = 404\ \mu \textrm{e}V$, gives rise to the ratio $2\Delta(0)/k_{B}T_c = 4.08$, indicating the strong coupling nature in CeCoIn$_5$ in agreement with other experiments.\cite{petrovic01}
However, we note that $\Gamma$ decreases with increasing temperature, which is unphysical and in contrast to usual observations\cite{dynes78,incgamma} that $\Gamma$ increases with increasing temperature. 
We attribute this behavior to the \textit{failure of the $s$-wave BTK model} to account for our data.

In the case of the $d$-wave BTK model applied to a (001) PCJ, it is not possible to distinguish between $d_{x^2-y^2}$ and $d_{xy}$ symmetry since the junction normal is along the $c$-axis.
The best fit curve for the 400 mK data using the $d$-wave model is displayed as a solid line in Fig.~\ref{fig:001-400mK-n-zbc-fit}(a), together with the $s$-wave fit curve.
Here, we point out that the shallow dip seen around --1.2 mV in the data is not an intrinsic feature indicative of the local heating effect,\cite{naidyuk98,gloos00} but an artifact caused in the normalization process due to an imperfect match of the background conductances at 400 mK and 2.6 K.
The best fit values of the fitting parameters are, $Z_{eff}=0.365$, $\Gamma=218\ \mu \textrm{e}V$,
and $\Delta=460\ \mu \textrm{e}V$, which gives the ratio $2\Delta/k_{B}T_c = 4.64$,
again implying the strong coupling nature.\cite{petrovic01}
As shown in Fig.~\ref{fig:001-400mK-n-zbc-fit}(a), the $d$-wave model gives a slightly better fit than the $s$-wave model, showing less deviation (albeit still substantial) above the gap edge and reproducing a slight dip-peak feature near zero-bias. However, we cannot fit the conductance data over the whole temperature range without using $\Gamma$ which shows an unphysical temperature dependence, i.e., decreasing with increasing temperature, similarly in the $s$-wave fit in Fig.~\ref{fig:001s-wavefit}(b).
We interpret these results as a failure of the $d$-wave EBTK model to explain the reduction in both the energy and the ESGC in our conductance curves.

The ZBC would be least affected by any local heating effect.\cite{naidyuk98,gloos00}
Figure~\ref{fig:001-400mK-n-zbc-fit}(b) shows that the ZBC vs. temperature data can be equally fit using both the $s$- and the $d$-wave models. The $s$-wave fit curve is obtained using $\Delta(0)=349\  \mu \textrm{e}V$
and $Z_{eff}=0.346$. However, again, the parameter $\Gamma$ is required to have an unphysical temperature dependence, $\Gamma(t) = 0.86\Delta(0)(1-t^3/3)$, where $t=T/T_c$, as plotted in Fig.~\ref{fig:001s-wavefit}(b).
Combined together, these observations strongly indicate a breakdown of the $s$-wave BTK model in a (001) PCJ of CeCoIn$_5$. This is not unexpected since most reports on the order parameter symmetry of this material support the existence of line nodes.\cite{movshovich01,izawa01,eskildsen03,aoki04}
Meanwhile, for the $d$-wave model, the fitting parameters are $Z_{eff}=0.365$, 
$\Delta(T)=2.35k_{B}T_c\tanh(2.06\sqrt{T_c/T-1})$, and $\Gamma = 218\ \mu \textrm{e}V$, constant over the temperature range.
This constancy of $\Gamma$ in the $d$-wave fit is not unphysical, \textit{in contrast to the $s$-wave fit}. Thus, we argue that \textit{the $d$-wave is a more likely pairing symmetry} rather than the $s$-wave, consistent with the literature.\cite{thompson03,movshovich01,izawa01,eskildsen03,aoki04}

As seen above, however, the $d$-wave BTK model does not fully account for our data taken over the full range of temperature below $T_c$. We have investigated possible origins for this failure.
First, we note that our observation of AR-like conductance spectra is consistent with the PCS results on other N/HFS point contacts.\cite{goll93,wilde94,goll95,wilde96,naidyuk96PhysicaB,obermair98}
This is also in agreement with Deutscher and Nozi{\'e}res' arguments \cite{deutscher94},
in the sense that these N/HFS point contacts do not exhibit tunneling-like conductance features.
In addition, the ESGC is commonly observed to be suppressed heavily by an order of magnitude, compared to that in N/conventional superconductor contacts.
Our point contact spectra, although consistent with the data on other N/HFS contacts, constitute the cleanest data set over a wide temperature range. Since the ESGC is observed to accompany the superconducting transition, proximity\cite{pe} and pressure effects\cite{gloos95,gloos00} are ruled out as an origin of the suppressed AR. Likewise, the ballistic nature of the contact excludes the local heating effect\cite{naidyuk98,gloos00} and the dominant Maxwell resistance.\cite{gloos96}
As a result, we claim that there must be intrinsic origins causing AR-induced conductance in N/HFS contacts to be reduced severely.

Golubov and Tafuri\cite{golubov00} considered a breakdown of the Andreev approximation (retro-reflectivity) when $\Delta/E_F$ ($E_F$ is the Fermi energy) is not negligible and/or the electrodes have layered structures as in the HTSs.
Considering the small $E_F$ of 15 K for CeCoIn$_5$,\cite{kasahara05} the non-retro-reflectivity may contribute to the reduced ESGC.
Mortensen \textit{et al}.\cite{mortensen99} considered mismatches in Fermi velocities and momenta and showed that the ESGC can be reduced in the case of large disparity.
However, in their calculation, what is reduced is the SGC normalized by the normal state conductance instead of high-bias conductance ($V \gg \Delta/e$).
As shown in Fig.~\ref{fig:001-400mK-n-zbc-fit}(a), the measured conductance above the gap edge is reduced compared to the calculated one. This may imply that the energy scale, which can be represented by $\Delta$, is also reduced in CeCoIn$_5$ as well as the ESGC. Therefore, it is unlikely that the measured data can be accounted for with only mismatches in Fermi surface parameters.\cite{golubov00,mortensen99}
Looking at the two-fluid model,\cite{nakatsuji04} one can formulate a model in which the effective barrier strength has different values for electrons and holes in each of the two fluids.\cite{golubov00} This kind of theory would naturally account for the asymmetry in the background conductance.
On the other hand, it is found that conductance calculations, based on the EBTK models and taking into account spatial variations of the effective mass and the order parameter, just give rise to usual BTK conductances with proper scalings of the parameters.\cite{lukic-elenewski}
Anders and Gloos\cite{anders97} proposed that $\Gamma$ in HFSs may be strongly dependent on the energy. Their calculated conductance curves for N/HFS point contacts based on the Green function formalism seems to be consistent with our data in the sense that both the energy gap and the SGC are reduced, compared to the BTK conductance for usual N/S point contacts. Nevertheless, this theory has a drawback that it is not easy to track down the physical mechanism clearly. More rigorous and detailed investigations of this model are necessary.
One of other issues to be addressed is about the relevant time scale in regard to the QP broadening factor $\Gamma$: it has not been studied well for a multi-particle AR process, whereas plenty of investigations have been made for a single particle tunneling process.\cite{dynes78} The directionality of charge transport due to the quasi-two dimensional nature of the Fermi surfaces\cite{settai01} and the effect of non-Fermi liquid nature of CeCoIn$_5$\cite{sidorov02} should also be taken into consideration.

\subsection{Conductance of (110) Point-Contact Junction}

\begin{figure}[b]
\begin{center}
\includegraphics{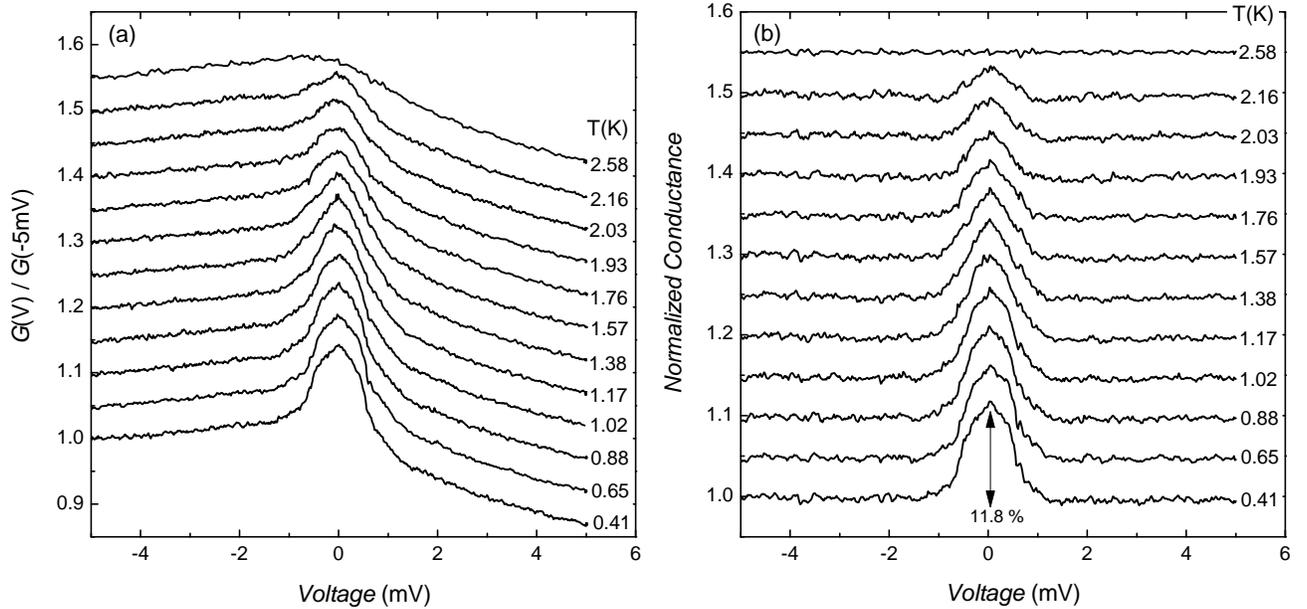}
\end{center}
\caption{Conductance vs. voltage data of a Au/CeCoIn$_5$ (110) point-contact junction. Curves are shifted vertically by 0.05 for clarity. (a) Conductance data are normalized to the conductance at --5 mV. The asymmetry in the background conductance remains constant over the measured temperature range, similarly to the (001) point-contact junction.
(b) Conductance curves are symmetrized by dividing out the data in (a) by the one at at 2.58 K. The conductance enhancement near zero-bias, 11.8 \% at 410 mK, is due to Andreev Bound States.}
\label{fig:110temp-dep}
\end{figure}

Figure~\ref{fig:110temp-dep}(a) shows conductance spectra taken from a (110) PCJ. It is seen that the asymmetric background conductance remains almost the same over the temperature range from 2.58 K to 410 mK, which is also observed in the (001) PCJ. Therefore, the asymmetry in the background conductance seems to be an intrinsic feature of the Au/CeCoIn$_5$ point contacts. Below $T_c$, an enhancement of the conductance appears near zero-bias.
Since $R_N$ of this PCJ is $\sim 4.7\ \Omega$, the contact size is about half of that for the (001) PCJ, ensuring the ballistic nature of the point contact. 
Similarly to the case of (001) PCJ, the conductance curves are symmetrized by dividing out the conductance data by the one at 2.58 K, as plotted in Fig.~\ref{fig:110temp-dep}(b). The conductance curves at lower temperatures exhibit a change of the slope in the sub-gap region, which resembles the curve for small $Z_{eff}$ in Fig.~\ref{fig:eBTK-calc}(b). It is remarkable that the cusp-like conductance feature near zero-bias persists up to the highest measured temperature below $T_c$. 

\begin{figure}[t]
\begin{center}
\includegraphics{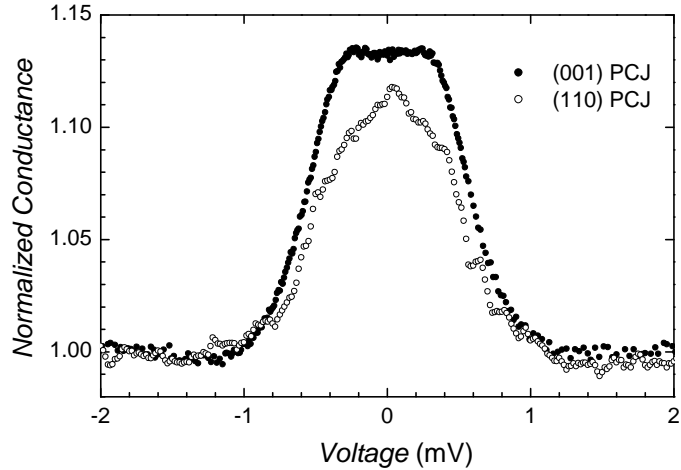}
\end{center}
\caption{Comparative plots of the symmetrized conductance data for the (001) and (110) point-contact junctions at 400 mK and 410 mK, respectively. They are consistent in the sense that they have similar values for the zero-bias conductance and the energy scale for the conductance enhancement. The shape of the conductance curve in the sub-gap region is strikingly different from each other. The cusp-like shape at zero-bias and the slope change near $\sim 400\ \mu \textrm{e}V$ indicate that the (110) point-contact junction is normal to the nodal direction, consistent with the $d_{x^2-y^2}$ symmetry.}
\label{fig:001vs110}
\end{figure}

The conductance data taken at the lowest temperatures are plotted together in Fig.~\ref{fig:001vs110}. They are similar to each other in terms of the ZBC enhancement(13.3 \% vs. 11.8 \%) and the energy scale for the conductance enhancement ($\sim 1\ \ m\textrm{e}V$).
However, the shape of the conductance curve is strikingly different near zero-bias: flat vs. cusp-like for the (001) and (110) PCJs, respectively. These different shapes are consistent with the calculated curves in Fig.~\ref{fig:eBTK-calc}, where clearly different behaviors of the conductance curve are observed with varying $Z_{eff}$. It is quite natural that the Au/CeCoIn$_5$ point contact may have non-zero $Z_{eff}$ because of the disparate Fermi velocities, even after the Deutscher-Nozi{\`e}res theory\cite{deutscher94} is taken into consideration. If the (110) PCJ were formed with $\alpha=0$, the conductance would then show a dip or flat structure near zero-bias, as in Fig.~\ref{fig:eBTK-calc}(a). The cusp-like structure in the conductance curve indicates that $\alpha \neq 0$. The data are consistent with $\alpha=\pi/4$, as shown in Fig.~\ref{fig:eBTK-calc}(b). This is consistent with the (110) PCJ being formed along the nodal direction of the order parameter, which supports the $d_{x^2-y^2}$-wave symmetry.\cite{izawa01,eskildsen03} As demonstrated in Fig.~\ref{fig:eBTK-calc}(b), the ZBCP of a PCJ with $\alpha = \pi/4$ increases rapidly with increasing $Z_{eff}$. The magnitude of the ZBCP is comparable to that of the (001) PCJ, implying that $Z_{eff}$ is non-zero but small, maybe around 0.3, comparable to that for the (001) PCJ.

The ZBCP in HFS junctions has been rarely observed, in contrast to HTS junctions.\cite{covington97} This is partly because the techniques for thin film growth and junction fabrication are not well established. It is remarkable that our conductance spectra from the (110) PCJ show clear signatures of $d_{x^2-y^2}$-wave symmetry, thereby addressing the controversial issue\cite{izawa01,aoki04} on the location of line nodes. However, as in the HTSs, high quality tunnel junctions are desirable to provide more conclusive evidence for the ABS-induced ZBCP. A complete and quantitative analysis on the conductance data from the (110) PCJ is deferred until a theoretical model is set up so that the mechanism for the suppressed AR, a common observation from both (001) and (110) PCJs, can be understood.

\section{CONCLUSIONS}
\label{sect:conclu}

We have obtained conductance spectra from ballistic Au/CeCoIn$_5$ point-contact junctions with the normal along the $c$-axis and $ab$-plane directions, over wide temperature ranges. Our data constitute the cleanest sets ever reported for N/HFS point contacts. The data obtained from both junctions are consistent with each other in terms of the asymmetry in the background conductance, the magnitude of the ZBC, and the energy scale. Quantitative analyses of the conductance spectra for the (001) point-contact junction show that existing models cannot adequately describe the particle-hole Andreev conversion process at this interface. The gap energy extracted from the fit to the conductance curve at the lowest temperature implies the strong coupling nature. The temperature-dependence of a single point, the zero-bias conductance, is consistent with the $d$-wave order parameter symmetry. Both conclusions are consistent with the literature for CeCoIn$_5$. We propose that systematic corrections to the BTK model that go beyond the breakdown of the Andreev approximation and the re-normalized Fermi momenta may provide a framework for our future understanding of Andreev reflection at the N/HFS interface. Qualitative analysis of the conductance data for the (110) point-contact junction show clear signatures for the $d_{x^2-y^2}$-wave pairing symmetry. Our data provide the first spectroscopic evidence for the order parameter symmetry and the orientation of line nodes in CeCoIn$_5$.

\acknowledgments       

We are grateful to A. J. Leggett, D. Pines, V. Lukic, and J. Elenewski for fruitful discussions 
and to B. F. Wilken, A. N. Thaler, P. J. Hentges, K. Parkinson, and W. L. Feldmann for experimental help.
This work was supported by the U.S. Department of Energy Award No. DEFG02-91ER45439,
through the Frederick Seitz Materials Research Laboratory and the Center for Microanalysis of Materials at the University of Illinois at Urbana-Champaign. 


\end{document}